\documentstyle[aps,prb]{revtex}

\begin{document}
\title{On the free energy of ionic hydration}
\author{Gerhard Hummer,$^{*ab}$ Lawrence R. Pratt,$^c$ and
Angel E. Garc\'{\i}a$^a$}
\address{}
\date{April 6, 1995}
\maketitle
\begin{center}
{\em The Journal of Physical Chemistry} (in press, 1995)
\end{center}
\begin{abstract}
The hydration free energies of ions exhibit an approximately quadratic
dependence on the ionic charge, as predicted by the Born model. We analyze this
behavior using second-order perturbation theory. This provides effective
methods to calculating free energies from equilibrium computer simulations. The
average and the fluctuation of the electrostatic potential at charge sites
appear as the first coefficients in a Taylor expansion of the free energy of
charging. Combining the data from different charge states ({\em e.g.}, charged
and uncharged) allows calculation of free-energy profiles as a function of the
ionic charge. The first two Taylor coefficients of the free-energy profiles can
be computed accurately from equilibrium simulations; but they are affected by a
strong system-size dependence. We apply corrections for these finite-size
effects by using Ewald lattice summation and adding the self-interactions
consistently. An analogous procedure is used for reaction-field
electrostatics. Results are presented for a model ion with methane-like
Lennard-Jones parameters in SPC water. We find two very closely quadratic
regimes with different parameters for positive and negative ions. We also
studied the hydration free energy of potassium, calcium, fluoride, chloride,
and bromide ions. We find negative ions to be solvated more strongly (as
measured by hydration free energies) compared to positive ions of equal size,
in agreement with experimental data. We ascribe this preference of negative
ions to their strong interactions with water hydrogens, which can penetrate the
ionic van der Waals shell without direct energetic penalty in the models
used. In addition, we consistently find a positive electrostatic potential at
the center of uncharged Lennard-Jones particles in water, which also favors
negative ions. Regarding the effects of a finite system size, we show that even
using only 16 water molecules it is possible to calculate accurately the
hydration free energy of sodium if self-interactions are considered.
\end{abstract}
\vfill
\hrule\vspace*{0.5\baselineskip}\noindent
$^a$ Address for Correspondence:
Theoretical Biology and Biophysics Group T-10, MS K710,
Los Alamos National Laboratory, Los Alamos, New Mexico 87545\\
FAX: (505) 665-3493; Phone: (505) 665-1923;
E-mail: hummer@t10.lanl.gov\\[0.5\baselineskip]
$^b$ Center for Nonlinear Studies,
Los Alamos National Laboratory, MS B258, Los Alamos, New Mexico
87545\\[0.5\baselineskip]
$^c$ Theoretical Chemistry and Molecular Physics Group T-12, MS B268,
Los Alamos National Laboratory, Los Alamos, New Mexico 87545

\section{Introduction}
A {\em quadratic} dependence on the ionic charge of the electrostatic
free energy of solvation of a simple ion in aqueous solution is about
the simplest reasonable possibility for that behavior. The Born model
predicts that quadratic dependence.\cite{Born:20} Several computer
simulation calculations have shown that it is approximately correct
for the simplest monovalent ions in
water.\cite{Jayaram:89,Levy:91,Smith:94} Theoretical simplifications
have been advanced to take advantage of such
behavior.\cite{Levy:91,Pratt:94:a,Tawa:94:a,Tawa:95:a}

If that quadratic behavior were correct with sufficient accuracy, it
would indeed permit important simplifications of the difficult task of
molecular calculations of solvation free energies owing to electrostatic
interactions in complex solutions. The theoretical simplifications
identified on that basis can be viewed either as perturbation theory
through second order in the electrostatic interactions, or as a
Gaussian modeling of certain thermal fluctuations of those
interactions. Adopting either view, these methods would have wide
applicability and great simplicity. The question of the accuracy of
the quadratic dependence on charge of the free energy owing to
electrostatic interactions deserves to be raised for its own sake and
given a precise answer as general as possible.

This quadratic behavior is not a universal truth and previous
simulation calculations have given helpful information on the
circumstances where this quadratic dependence can be expected to
fail.\cite{Jayaram:89} However, previous simulation calculations are
sufficiently disparate that a high precision answer to the question of
the accuracy of second-order perturbation theory for the free energy
owing to electrostatic interactions is not available. The disparate
character of the available simulation results is largely caused by a
lack of uniformity with respect to the treatment of finite-system-size
effects on electrostatic interactions in aqueous solutions. It is not
atypical for a finite-system-size correction and the electrostatic
solvation free energy to be of similar size.

In contrast to the role of computer experiments in answering this
question, laboratory experiments have been useful mostly for framing the
question.\cite{Honig:93,Warshel:94:b,Marcus:94:a} The difficulty of
using laboratory experiments for the present purpose resides in
our inability to extract generally an electrostatic contribution from
contributions of the other interactions present.

Because of these points, this work calculates the free energy owing to
electrostatic interactions of simple, spherical ions in water by
Monte Carlo methods and gives particular attention to the
methodological issue of correction for finite system size. The
molecular models used are simple but they have been widely tested.
Because the goal of this work is to address the question of quadratic
dependence on charge of the electrostatic solvation free energy, these
models are sufficiently realistic for the present purposes. However,
we will compare our computed free energies with experimental results
and thus provide information on how these models might be simply
improved for prediction of electrostatic free energies.

Before proceeding with the technical developments it is worthwhile to
give some discussion of the idea for the present treatment of
system-size effects on solvation free energies of ions. There is no
generally valid recipe that allows a determination of the effects of a
finite system size on the calculated physical quantity in computer
simulations. What must generally be done is to analyze the observed
size dependence empirically. If, as is the case for Coulomb
interactions of long range, different procedures are available, then
we should expect consistent thermodynamic limiting
($N\rightarrow\infty$) results for different methods of treating the
finite-size system. It is well understood that certain quantities
involving integrals over the whole sample, such as the dipole-moment
fluctuations, depend intrinsically on exterior conditions or
constraints.\cite{Neumann:83:b} Those conditions must then be properly
understood theoretically.

For the present problem involving the interactions and associated
thermodynamics of an ion immersed in a dielectric liquid, a reasonable
view is the following: Treatment of electrostatic interactions in a
truly periodic format, {\em e.g.\/}, by Ewald procedures, is consistent
with the periodic boundary conditions that are nearly inevitable for
other reasons. In periodic boundary conditions the interactions at the
longest range that {\em must\/} be taken seriously occur at an
appreciable fraction of the distance to the surface of the simulation
cell. For typical simulated system sizes, ionic interactions at that
longest range are large. Treatment of electrostatic interactions in a
truly periodic format thoroughly tempers those large interactions. But
a mathematical price for true periodicity of electrostatic interactions
is a ``self-interaction'' associated with interactions with images and a
uniform neutralizing charge background. For neutral systems this
self-interaction can be sometimes ignored. For non-neutral systems such
as those studied here there may be practical advantages of consistency
obtained for explicit consideration of the self-interaction. We will
account for these self-interactions explicitly in the calculations
below.

This argument permits treatments of the ionic interactions other than
Ewald summation. In fact, the work below tests a generalized
reaction-field (GRF) method and also finds that consistent results can
be obtained if self-interactions are treated on a similar basis.

\section{Theoretical methods}
\subsection{Calculation of the free energy of charging}
The various methods to compute free energies using computer
simulations have been reviewed
extensively.\cite{Frenkel:86,Allen:87,Levesque:92} We start here from
the potential distribution theorem for the excess chemical potential
$\mu^{ex}$,\cite{Widom:82}
\begin{eqnarray}
\mu^{ex}(q_1)-\mu^{ex}(q_0)&=&-k_BT\ln
\left\langle {\exp\left\{-\beta [u(q_1)-u(q_0)]\right\}}
\right\rangle_{q_0}~,
\label{eq:dF}
\end{eqnarray}
where $q_0$ and $q_1$ are the two charge states and $\beta=1/k_{\rm
B}T$; $\langle \ldots \rangle_q$ denotes a thermal configuration-space
average in the charge-state $q$; and $u(q)$ is the
configuration-dependent interaction energy of the ion in charge-state
$q$ with the solution. Apart from finite-size corrections to be
discussed later, $u(q)$ is given by $q\phi({\bf r})$, where $\phi({\bf
r})$ is the electrostatic potential at the charge position ${\bf r}$.

We next analyze eq~\ref{eq:dF} utilizing a cumulant
expansion\cite{Kubo:62} with respect to $\beta$,
\begin{eqnarray}
\left\langle \exp\left( -\beta \Delta u \right)
\right\rangle_{q_0} & = & \exp\left[ \sum_{n=0}^{\infty} (-\beta)^n
\frac{C_n}{n!} \right]~,
\label{eq:dFC}
\end{eqnarray}
where $\Delta u = u(q_1)-u(q_0)$. This defines the cumulants $C_n$ of
order $n=0,1,2$ as
\begin{mathletters}
\begin{eqnarray}
C_0 & = & 0\\
C_1 & = & \left\langle \Delta u \right\rangle_{q_0}\\
C_2 & = & \left\langle \left( \Delta u - \left\langle
\Delta u \right\rangle_{q_0} \right)^2 \right\rangle_{q_0}~.
\end{eqnarray}
\end{mathletters}
We can interpret eq~\ref{eq:dFC} as a Taylor expansion in $\Delta q =
q_1 -q_0$ if we set $\Delta u = \Delta q \phi + ( q_1^2 - q_0^2 ) \xi /
2 $, where $\xi$ accounts for finite-size effects as a
``self-interaction'' to be discussed further below,
\begin{eqnarray}
\Delta \mu^{ex} & = & \Delta q \left( \left\langle \phi
\right\rangle_{q_0} + q_0 \xi \right) - \frac{\beta}{2} \Delta q^2
\left[ \left\langle \left( \phi - \left\langle \phi \right\rangle_{q_0}
\right)^2 \right\rangle_{q_0} - \frac{\xi}{\beta} \right]+\cdots~,
\label{eq:dmuC}
\end{eqnarray}
where $\Delta \mu^{ex}=\mu^{ex}(q_1)-\mu^{ex}(q_0)$.
The mean and the fluctuation of the electrostatic potential at the
charge site $q$ (corrected for finite-size effects) yield the
derivatives of the free energy with respect to $\Delta q$. The
information about the derivatives can therefore be extracted from
equilibrium simulations. In principle, higher-order cumulants could be
used to obtain information about the other Taylor coefficients.
However, as was observed by Smith and van Gunsteren,\cite{Smith:94}
higher-order cumulants are increasingly difficult to extract from
computer simulations of limited duration.

Therefore, we will evaluate $C_1$ and $C_2$ at few discrete charge
states and combine this information about the derivatives, either by
constructing an interpolating polynomial or by using a $\chi^2$ fit to
a polynomial expression (or any other functional form) for the free
energy as a function of $\Delta q$. The $\chi^2$ fit minimizes the mean
square deviation of the observed data with respect to the coefficients
$\{a_k\}$ of the fitting function $\Delta\mu^{ex}(q;\{a_k\})$,
\begin{eqnarray}
\chi^2 & = & \sum_{i=1}^{n} \left\{
\left[ \frac{\Delta \dot \mu^{ex}(q_i;\{a_k\})
- \Delta \dot \mu^{ex}_{\rm obs}(q_i)}{\sigma_i} \right]^2 +
\left[ \frac{\Delta \ddot \mu^{ex}(q_i;\{a_k\})
- \Delta \ddot \mu^{ex}_{\rm obs}(q_i)}{\rho_i} \right]^2 \right\},
\label{eq:chi2}
\end{eqnarray}
where $\sigma_i$ and $\rho_i$ are the estimated errors (standard
deviations) of the observed first and second derivatives
$\dot \mu^{ex}_{\rm obs}$ and $\ddot \mu^{ex}_{\rm obs}$ at charge-state
$q_i$.

\subsection{Long-range Coulomb interactions and finite-size effects}
To minimize finite-size effects on energetic properties of Coulombic
systems, we adopt the following strategy:\cite{Hummer:93} We use
lattice summation for calculating the electrostatic interactions to
account for the periodic boundary conditions employed in the computer
simulations; and we consistently include the self-interactions arising from
lattice summation. We point out that aside from formal consistency
the numerical results can motivate this approach by demonstrating in a
finite-size analysis that the deviations from the thermodynamic limit
($N\rightarrow\infty$) are small.

The Coulomb energy of a periodically replicated system of charges $q_i$
at positions ${\bf r}_i$ ($i=1,\ldots,N$) can be expressed as
\begin{eqnarray}
U & = & \sum_{1\leq i<j\leq N} q_i q_j \varphi_{\rm EW}({\bf
r}_{ij}) + \frac{1}{2} \sum_{1\leq i\leq N} q_i^2 \xi_{\rm EW}~,
\label{eq:EwaldTot}
\end{eqnarray}
where ${\bf r}_{ij}={\bf r}_j-{\bf r}_i+{\bf n}$, with the lattice
vector ${\bf n}$ chosen such that ${\bf r}_{ij}$ is a vector in the
unit cell. The effective, position-dependent potential $\varphi_{\rm
EW}({\bf r})$ is obtained by lattice summation using Ewald's
method,\cite{Allen:87,Ewald:21,deLeeuw:80:a}
\begin{eqnarray}
\varphi_{\rm EW}({\bf r}) & = &
\sum_{\bf n}\frac{{\rm erfc}(\eta|{\bf
r}+{\bf n}|)} {|{\bf r}+{\bf n}|} + \sum_{{\bf k}\neq 0}
\frac{4\pi}{V k^2} \exp \left( -\frac{k^2}{4\eta^2} +
i{\bf k}\cdot{\bf r}\right) -\frac{\pi}{V \eta^2}~,
\label{eq:Ewald}
\end{eqnarray}
where $V$ is the volume of the box, erfc is the complementary error
function, and $k=|{\bf k}|$. The two lattice sums extend over real- and
Fourier-space lattice-vectors ${\bf n}$ and ${\bf k}$,
respectively.

The self term $\xi_{\rm EW}= \lim_{{\bf r}\rightarrow 0} [
\varphi_{\rm EW}({\bf r}) -1/r]$ is the Wigner
potential:\cite{deLeeuw:86,Cichocki:89,Nijboer:88} Using Green's
theorem and $\Delta (1/r) = -4\pi\delta({\bf r})$, we find
\begin{eqnarray}
\xi_{\rm EW} = \lim_{{\bf r}\rightarrow 0}\left[
\varphi_{\rm EW}({\bf r}) - \frac{1}{r} \right] & = &
-\frac{1}{4\pi} \lim_{\epsilon\rightarrow 0}
\int_{|{\bf r}|>\epsilon}
d{\bf r} \;\frac{1}{r}\;\Delta\varphi_{\rm EW}({\bf r})~.
\label{eq:xiew}
\end{eqnarray}
The integration region is infinite and includes all background charge
and lattice image charges,
\begin{eqnarray}
\Delta \varphi_{\rm EW}({\bf r}) &=& -4 \pi \sum_{\bf n}
\left[ \delta({\bf r}-{\bf n}) - \frac{1}{V}\right]~.
\label{eq:LapEw}
\end{eqnarray}
Eqs~\ref{eq:xiew} and \ref{eq:LapEw} establish that $\xi_{\rm EW}$
is the electrostatic potential in a Wigner lattice at a charge site owing
to the lattice images and the neutralizing background.
For Ewald summation in a cubic lattice the self term is $\xi_{\rm
EW}=-2.837297/L$,\cite{Cichocki:89,Nijboer:88,Placzek:51} where $L$ is
the length of the cube.

It will be interesting to remember that $\xi_{\rm EW}$ can also be
expressed in terms of quantities associated with the primitive
simulation cell
\begin{eqnarray}
\xi_{\rm EW} & = & -\frac{1}{4\pi}
\lim_{\epsilon\rightarrow 0} \int_{V:|{\bf r}|>\epsilon}
d{\bf r} \;\frac{1}{r}\;\Delta\varphi_{\rm EW}({\bf r})
-\frac{1}{4\pi}\int_{\partial V} d^2r\; \varphi_{\rm EW}({\bf r})\;
\hat {\bf n} \cdot \nabla \left( \frac{1}{r}\right).
\label{eq:be}
\end{eqnarray}
The first term on the right is explicitly the interaction with the
background density in the primitive simulation cell. The second term
on the right is an integral over the surface of the primitive
simulation cell. It describes electrostatic interactions of the
central unit charge with a dipolar surface distribution $ \varphi_{\rm
EW}({\bf r})\;\hat {\bf n}$, where $\hat {\bf n}$ is the surface
normal pointing outwards.

Eq~\ref{eq:EwaldTot} can also be used for a non-neutral system since
charges are implicitly compensated by a homogeneous background in the
Ewald formulation. This results in an expression for the energy
difference $\Delta u$ between two configurations with different charge-states
$q_0$ and $q_1$ of an ion at position ${\bf r}$,
\begin{eqnarray}
\Delta u & = & \Delta q \; \varphi_{\rm EW}({\bf r})
+ \frac{1}{2} \; \xi_{\rm EW} \; ( q_1^2 - q_0^2 )~.
\end{eqnarray}
In the following, we will use this expression containing a self-interaction
which is quadratic in the charge to calculate the free energy of
charging; {\em i.e.}, we assume that the self-interaction accounts for the
finite-size corrections.\cite{Hummer:92:b}

In our calculations, we will also use a generalized reaction
field (GRF).\cite{Hummer:94:a,Hummer:94:e} The GRF
Coulomb interaction depends only on the distance $r$ of
the charges and has a cutoff-distance $r_c$,
\begin{eqnarray}
\varphi_{\rm GRF}(r) & = &
\frac{1}{r} \; p(r/r_c) \; \Theta(r_c-r) + C \label{eq:GRF}~.
\end{eqnarray}
$\Theta$ is the Heaviside unit-step function; $p(x)$ is a screening
polynomial:
\begin{eqnarray}
p(x) & = & ( 1 - x )^4 ( 1 +8x/5 +2x^2/5 )~.
\label{eq:polyGRF}
\end{eqnarray}
By analogy with Ewald summation, we define the self term for the GRF as
the potential at the charge site, $\xi_{\rm GRF}= \lim_{{\bf
r}\rightarrow 0} [ \varphi_{\rm GRF}({\bf r}) -1/r]$.
The total energy of
neutral systems, if defined as in eq~\ref{eq:EwaldTot}, is
independent of the constant $C$. However, in non-neutral systems $C$
affects the total energy. We define $C$ in analogy with the Ewald
potential, which satisfies\cite{Nijboer:88}
\begin{eqnarray}
\int_{V}d{\bf r} \; \varphi_{\rm EW}({\bf r}) = 0~,
\label{eq:phinorm}
\end{eqnarray}
such that the average potential in the cell vanishes. If
we require the normalization condition eq~\ref{eq:phinorm}
also for the GRF interaction, we obtain $C = -\pi r_c^2 / 5V$.
The GRF self term is $\xi_{\rm GRF}=-12/5r_c+C$. For $r_c=L/2$, the
normalization condition eq~\ref{eq:phinorm} accounts for only a small
additional correction yielding $\xi_{\rm GRF}=-24/5L-\pi/20L$.

It is interesting to make a connection with the correction method
proposed by Sloth and S{\o}rensen.\cite{Sloth:90} These authors use
the minimum-image Coulomb interaction. To eliminate the system-size
dependence in their calculation of chemical potentials of
restricted-primitive-model ions, they introduce a background
neutralizing the test-particle charge. This is done by adding a
constant $\xi_{1/r}$ to the bare Coulomb
potential,\cite{minimg}\nocite{Sorensen:91}
\begin{eqnarray}
\xi_{1/r}&=&-\frac{1}{V}\int_{V}d{\bf r}\;\frac{1}{r}~.
\end{eqnarray}
This corresponds to enforcing eq~\ref{eq:phinorm} and adding
a self term $\xi_{1/r}=\lim_{r\rightarrow 0}[\varphi(r)-1/r]$ for the
minimum-image interaction. $\xi_{1/r}$ is also precisely the first
term on the right side of eq~\ref{eq:be}. It accounts for a large
correction since $\xi_{\rm 1/r}\approx -2.38/L$.

\section{Computer simulations}
\label{sec:methods}
We calculated the free energy of charging ions in water using
Metropolis Monte Carlo simulations.\cite{Allen:87,Metropolis:53} The
systems comprise a single ion and $N$ water molecules. For water we
used the simple point charge (SPC) model.\cite{Berendsen:81} The
ion-water interactions were described by Coulomb and Lennard-Jones
(LJ) interactions. The Coulomb terms involve the partial charges of
oxygens and hydrogens on SPC water. The LJ interactions act only
between water oxygen and the ion. We studied the ions Na$^+$, K$^+$,
Ca$^{2+}$, F$^-$, Cl$^-$, and Br$^-$. The LJ parameters for these
ions were those of Straatsma and Berendsen.\cite{Straatsma:88} We also
studied the charging of a model ion Me with methane LJ parameters as
given by Jorgensen {\em et al.}\cite{Jorgensen:84} Lorentz-Berthelot
mixing rules\cite{Allen:87} were applied to obtain LJ parameters
between water and Me. The LJ parameters are compiled in
Table~\ref{tab:LJ}.

The charge interactions in the simulations were calculated using Ewald
lattice summation (eqs~\ref{eq:EwaldTot} and \ref{eq:Ewald}) and the
generalized reaction-field potential (eqs~\ref{eq:GRF} and
\ref{eq:polyGRF}). In both cases, the real-space interactions were
truncated on an atom basis using $L/2$ as cutoff and applying the
periodic boundary conditions on an atom basis. For the Ewald
Fourier-space calculation, a cutoff $k^2\leq 38(2\pi/L)^2$ was used
resulting in $2 \times 510$ ${\bf k}$ vectors. To correct the
background dielectric constant from infinity to $\epsilon_{\rm
RF}=65$, a term $2\pi{\bf M}^2/(2\epsilon_{\rm RF}+1)V$ was added to
the potential energy (in both Ewald and GRF calculations), where ${\bf
M}$ is the net dipole moment of the water molecules. The real-space
damping factor was set to $\eta=5.6/L$. Electrostatic potentials at
the ion sites were calculated using $\varphi_{\rm EW}$ and
$\varphi_{\rm GRF}$. The potentials were calculated after each pass
(one attempted move per particle) and stored for analysis. For each
system 100\,000 passes were used for averaging. Random configurations
or configurations of previous runs were used as initial structures and
always extensively equilibrated. The temperature was 298~K. The total
number density was $\rho=33.33$~nm$^{-3}$ in all simulations. Cubic
boxes were used as simulation cells with edges
$L=[(N+1)/\rho]^{1/3}$. The Monte Carlo move widths were chosen so
that an approximate acceptance ratio of 0.5 was obtained.

In addition, thermodynamic integration (TI) was used to calculate
directly the free energy of charging. Within 100\,000 Monte Carlo
passes, the charge of the ion was linearly changed from 0 to its full
magnitude ($\pm e,2e$, where $e$ is the elementary charge).
The free-energy changes were then
calculated as
\begin{eqnarray}
\mu^{ex}(q_1) - \mu^{ex}(q_0) & = &
(q_1 - q_0) n^{-1} \sum_{i=1}^{n} \phi_i +
\xi ( q_1^2 - q_0^2 ) / 2~,
\label{eq:TI}
\end{eqnarray}
where the sum extends over $n=100\,000$ Monte Carlo passes and the
last term is a finite-size correction. Eq~\ref{eq:TI}
approximates the exact expression
\begin{eqnarray}
\mu^{ex}(q_1) - \mu^{ex}(q_0) & = & \int_{q_0}^{q_1} dq\;
\left\langle \frac{\partial
u(q)}{\partial q} \right\rangle_{q}~.
\end{eqnarray}
TI was also performed using the reverse
path, {\em i.e.}, decreasing the charge to 0.

We also performed Monte Carlo simulations of ion-water clusters
comprising one ion and $N$ SPC water molecules ($4\leq N\leq 256$).
The starting structure was a random configuration with bulk density of
water in a cubic box around the ion. The cluster was equilibrated for
at least 50\,000 passes (with an acceptance rate of about 0.5) and
then averaged over 50\,000 passes at 298~K. We used the bare Coulomb
interaction $1/r$ and did not apply a distance cutoff. No periodic
boundary conditions were employed in the cluster simulations.

\section{Results and discussion}
\subsection{Charging of a methane-like Lennard-Jones particle}
The free energy of charging a methane-like LJ particle in SPC water was
determined from a series of simulations with $N=128$ and 256 water
molecules and with Ewald and GRF charge treatment. A range of charges
from $-e$ to $e$ was covered in steps of $0.25e$ ($N=128$) and
$0.5e$ ($N=256$). The results for the mean $m$ and the fluctuation
$f$ of the potential at the ion site (with and without finite-size
correction) are compiled in Table~\ref{tab:methane}. In the
calculations, the potential $\phi$ at the ion site ({\bf r}=0)
is defined as
\begin{eqnarray}
\phi & = & \sum_{i=1}^{N}\sum_{\alpha=1}^{3}q_{i_\alpha}
\varphi({\bf r}_{i_\alpha})~,
\end{eqnarray}
where the double sum extends over all water oxygen and hydrogen sites;
$\varphi$ is either $\varphi_{\rm EW}$ or $\varphi_{\rm GRF}$. The mean
$m$ and the fluctuation $f$ are calculated from 100\,000 Monte Carlo
passes as
\begin{mathletters}
\begin{eqnarray}
m & = & e\langle\phi\rangle~,\\
f & = & \beta e^2 \langle ( \phi-\langle\phi\rangle )^2 \rangle~.
\end{eqnarray}
\end{mathletters}
The corrected values for mean and fluctuation are defined as
$m_c=m+qe\xi$ and $f_c=f-e^2\xi$. The Taylor expansion of the free
energy of charging around a charge-state $q$ assumes the following
form:
\begin{eqnarray}
\Delta \mu^{ex} & = & \left( { \Delta q \over e} \right) m_c
- {1 \over 2} \left( { \Delta q \over e} \right)^2 f_c +
\cdots~.
\end{eqnarray}
{}From the results of Table~\ref{tab:methane} we see that the
finite-size corrections are of similar magnitude as the uncorrected
results $m$ and $f$. The uncorrected results of the different methods
and system sizes are widely spread. If however the finite-size
corrections are applied, we obtain consistent results for all methods
and system sizes over the range of ion charges considered. With
estimated errors (one standard deviation, as calculated from block
averages) of 4.0 and 30 kJ~mol$^{-1}$ for $m_c$ and $f_c$, we find
data of different methods within two standard deviations from each
other. The only exception is the fluctuation $f_c$ for $q=-e$, where
the two extreme values (Ewald, $N=128$ and 256) differ by about three
standard deviations. In the following, we will restrict the
discussion to the corrected values $m_c$ and $f_c$.

Figure~\ref{fig:histo} shows the probability distribution $P(e\phi)$
of the electrostatic energy $e\phi$ for an ionic charge of $q=e$, as
calculated from histograms. The $P(e\phi)$ curves follow closely
Gaussian distributions with mean and variance calculated from the
$\phi$ data. This reflects the approximate validity of second-order
perturbation theory in the ionic charge. However, the $P(e\phi)$
curves for Ewald summation with $N=128$ and 256, as well as GRF
with $N=256$ water molecules differ widely, both in the peak position
and in the width. To illustrate the importance of the finite-size
correction, we included in Figure~\ref{fig:histo} the Gaussian
distributions corresponding to the corrected values $m_c$ and $f_c$
for mean and variance. The application of the finite-size corrections
brings the three curves to very close agreement, yielding results that
are approximately independent of system size and treatment of
electrostatic interactions.

To further illustrate the importance of the finite-size correction, we
calculated $\langle\phi\rangle$ from the pair correlations of the
Ewald-summation simulation with $N=256$ water molecules as
\begin{eqnarray}
\langle\phi\rangle(R) & = & 4\pi\rho_{\mbox{\scriptsize H$_2$O}}
\int_0^{R} dr\;r^2\;
\varphi(r) \left[q_{\rm O} g_{\rm IO}(r)+2q_{\rm H} g_{\rm IH}(r)
\right]~.
\label{eq:gint}
\end{eqnarray}
$\rho_{\mbox{\scriptsize H$_2$O}}$ is the water density, $q_{\rm O}$
and $q_{\rm H}$ are the oxygen and hydrogen charge, and $g_{\rm IO}$
and $g_{\rm IH}$ are the ion-oxygen and ion-hydrogen pair correlation
functions. Figure~\ref{fig:phi} shows the results for the charge-state
$q=-e$ of the ion Me as a function of the integration cutoff $R$
for the bare Coulomb potential $\varphi(r)=1/r$ and $\varphi_{\rm GRF}$
with $r_c=L/2$. In both cases we included the finite-size correction
as a constant. The integration of the $1/r$ interaction extended into
the corners of the cube, using the correct weights. As a reference,
the Ewald result is shown as a straight line. All three methods
converge to within about 1~kJ~mol$^{-1}$, which has to be compared
with the estimated statistical error of 4~kJ~mol$^{-1}$ of the data.
The integrated $1/r$ interaction shows strong oscillations and only in
the corners of the cube does it approach its final value. The GRF
interaction on the other hand contains a large self term and within
two oscillations reaches its limiting value.

This illustrates an important point regarding the correction of
finite-size effects in the calculation of charge-related quantities.
We achieve agreement between different methods of treating Coulomb
interactions (Ewald summation, reaction field, bare Coulomb
interaction) if we (i) normalize $\varphi$ according to
eq~\ref{eq:phinorm} and (ii) add a self term $\xi=\lim_{{\bf
r}\rightarrow 0}[\varphi({\bf r})-1/r]$ to the energy. Further
demonstrations of the validity of these finite-size corrections will be
given in the discussion of the results for sodium and fluoride ions
in SPC water.

Figure~\ref{fig:meth_m} shows $m_c$ as a function of the charge. We
observe two linear regimes with different characteristics for $q<0$
and $q\geq 0$. Linear behavior of $m_c$ on the whole range of $q$
would reflect validity of the second-order perturbation theory. It
would imply Gaussian statistics of $\phi$ and, correspondingly, that
the coefficients in the Taylor expansion of order three and higher
vanish. However, since we observe a transition in the linear behavior
between charges of $-0.25e$ and 0, the statistics are only
approximately Gaussian. We note that from the $\phi$ data of 100\,000
passes it proved impossible to extract reliable information about the
Taylor coefficients (cumulants) of order three and higher. The second
Taylor coefficient $f_c$ can however be extracted accurately.
Figure~\ref{fig:meth_f} shows $f_c$ as a function of $q/e$. Included
in Figure~\ref{fig:meth_f} as lines are the values of $f_c$ estimated
from the linear fits of $m_c$ for $q<0$ and $q\geq 0$.

We have fitted the $m_c$ and $f_c$ data by a model with two Gaussian
regimes. Included in Figures~\ref{fig:meth_m} and \ref{fig:meth_f}
is a $\chi^2$ fit of the whole set of derivative data (38 data points)
to
\begin{eqnarray}
\mu^{ex} (q)- \mu^{ex}(0) & = & (a_+q+b_+q^2) [1+\tanh(c+dq)]/2 +
(a_-q+b_-q^2) [1-\tanh(c+dq)]/2~,
\label{eq:tanh}
\end{eqnarray}
where $\chi^2$ is defined as in eq~\ref{eq:chi2} with parameters
$a_+$, $b_+$, $a_-$, $b_-$, $c$, and $d$. This model can nicely
reproduce the data. We find a transition at $q=c/d\approx -0.2e$
between the two regimes of approximately Gaussian behavior with a
quadratic $q$ dependence. We ascribe this transition to differences
in the structural organization of water molecules near negatively and
positively charged ions. A possible explanation for the observed
behavior is that for positive ions, the oxygen atom of water is
pointing towards the LJ particle. The strongly repulsive forces of the
$r^{-12}$ interaction prevent large fluctuations of $\phi$ because of
the restricted oxygen motions. The hydrogens are pointing away so
that rearranging them has only a comparably small effect on $\phi$.
For negative ions, the structures with one of the hydrogens pointing
towards the ion will dominate. Because of the symmetry between the
water hydrogens and the finite life time of the hydration shell,
transitions will occur which could explain the larger fluctuations in
the negative charge range.

Similarly, a transition to a different Gaussian behavior for
highly-charged positive ions was observed by Jayaram {\em et
al.}\cite{Jayaram:89}\ \ These authors studied the free energy of
charging of a sodium ion in the charge range 0 to $3e$. When
increasing the ion charge, a transition occurs to a more weakly
decreasing quadratic free-energy regime at a charge of about
$1.1e$. This transition has also been discussed by Figueirido {\em et
al.}\cite{Figueirido:94}

We also find a nonvanishing potential at the methane site even at zero
charge.\cite{Pratt:94:a} In a dipolar solvent,
$\langle\phi\rangle_{q=0}$ is zero because of charge-reversal
symmetry. However, if higher multipole moments are present on the
solvent molecule, this symmetry is not conserved. The asymmetry of
the charge distribution on the water molecule gives rise to a positive
potential for $q=0$; this is primarily caused by the hydrogens
penetrating the LJ sphere of the methane particle, since they do not
have a protecting repulsive shell in the model used. As a
consequence, there is a small charge region in which increasing the
charge {\em costs} free energy. A positive potential at the center of
an uncharged particle was also observed by Rick and Berne.\cite{Rick:94}

As a consequence of both the positive potential at zero charge and the
larger potential fluctuations for negative ions, negative ions are
more stably solvated compared to positive ions.
Table~\ref{tab:df_meth} compiles the free energies of charging as
calculated from fitted polynomials $p_n$ of degree $n$ to the
derivative data $m_c$ and $f_c$. Except for the simple
Gaussian model $p_2$, different fitting functions give consistent
results for the free energies of charging. For ions with charge $e$
and $-e$ we find $\Delta \mu^{ex}=-250$ and $-431$~kJ~mol$^{-1}$.
Interpreted within a Born model for the free energy,\cite{Born:20}
{\em i.e.},
\begin{eqnarray}
\Delta \mu^{ex}_{\rm Born} = -(1-1/\epsilon) q^2/2R~,
\label{eq:Born}
\end{eqnarray}
we obtain Born radii $R_+=0.27$~nm and $R_-=0.16$~nm. (A value of
$\epsilon=80$ is used for the dielectric constant, but this hardly
affects the results). The difference between $R_+$ and $R_-$ is
somewhat smaller if we use the actual coefficients of the $q^2$ term
in the free energy expansion, as obtained from eq~\ref{eq:tanh},
giving 0.23 and 0.16~nm for the Born radii of positive and negative
ions. We emphasize the model character of the interaction potentials
used in this study. A repulsive shell of the hydrogen atom might
reduce the free energy difference between positive and negative ions.
The favoring of negative ions however should persist.

The lower free energy of negative ions compared to positive ions of
equal size agrees with the experimental observations. The
hydration free-energy data compiled by Marcus\cite{Marcus:91:a} for
alkali metal and halide ions show a power-law dependence with respect to
the ion radius. Using these fitted curves, we find differences of 150
and 240~kJ~mol$^{-1}$ for the solvation free energy between negative
and positive ions of the size of potassium and sodium, respectively.
The LJ particle Me studied here has a van der Waals radius between
those of K$^+$ and Na$^+$. The calculated free energy required to go
from $-e$ to $+e$ is 180~kJ~mol$^{-1}$, which is indeed bracketed
by the experimental data.

The revised Born model by Latimer {\em et al.}\cite{Latimer:39} also
yields lower free energies for negative ions. For alkali and halide
ions, it uses effective Born radii $R=r_{\rm P}+\Delta$, where $r_{\rm
P}$ is the Pauling radius and $\Delta$ is 0.085 and 0.010~nm for cations
and anions. This smaller effective-radius correction for anions in
eq~\ref{eq:Born} results in considerably lower free energies of
negative ions compared to positive ions of equal size in agreement with
our calculations. The difference of the effective Born-radius
correction as defined by Latimer {\em et al.}\cite{Latimer:39} is
0.075~nm, which agrees with what we find for the Me ion.

The energetic differences in the hydration of positive and negative
ions go along with differences in the structural organization of water
molecules in the hydration shell. Figure~\ref{fig:gr} shows the
ion-water pair correlation functions for different ionic charges.
Going from $q=0$ to positive charges does not change the qualitative
properties of the ion-oxygen and ion-hydrogen correlation functions
$g_{\rm IO}$ and $g_{\rm IH}$. An increase of the ionic charge results
in a higher first peak. However, going from charge $q=0$ to negative
charges affects strongly the structural organization of the first
hydration shell. Already at $q=-0.5 e$, $g_{\rm IH}$ shows the buildup
of a second peak at about $r=0.2$~nm distance. This peak reaches a
value of almost 5 at $q=-e$, compared to $g_{\rm IH}$ essentially
being zero in this distance region for charge $q=0$. This strong
interaction of the negatively charged ion with the hydrogens of water
in turn affects the ion-oxygen correlation functions. Despite the
negative charge of both the ion and oxygen site, $g_{\rm IO}$ has a
first peak with a height of about 5 for $q=-e$ compared to only 3 for
$q=e$. The strong charge repulsion between water oxygens and the ion with
$q=-e$ is overcome by a large attraction caused by a water hydrogen
pointing towards the ion and penetrating the ionic van der Waals shell
without energetic penalty.

\subsection{Free energy of charging of the ions Na$^+$, K$^+$,
Ca$^{2+}$, F$^-$, Cl$^-$, and Br$^-$} Using the LJ parameters of
Straatsma and Berendsen\cite{Straatsma:88} (see Table~\ref{tab:LJ}),
we computed solvation free energies of ions representing Na$^+$,
K$^+$, Ca$^{2+}$, F$^-$, Cl$^-$, and Br$^-$. Again, we emphasize the
model character of this study. Its purpose is not to provide accurate
theoretical values for the free energies but rather to characterize
the theory. We can expect to obtain accurate values only after
considerable improvement of the currently rather crude descriptions of
the interaction potentials used here and similarly in most other
studies. Some of that work has indeed been guided by using free
energies of hydration.\cite{Aqvist:90,Marrone:93} However,
controversies about certain technical aspects, primarily regarding the
correct treatment of long-range interactions, need to be resolved to
obtain conclusive results.\cite{Aqvist:94,Marrone:94}

We extensively studied the solvation free energy of the sodium cation
using the model described in section~\ref{sec:methods}. Monte Carlo
simulations using $N=128$ water molecules were carried out for charges
0, $0.5e$, and $1.0e$ to calculate the mean $m_c$ and the fluctuation $f_c$
of the electrostatic potential $\phi$ at the ion site. As in the
previous calculations, 100\,000 passes were used for averaging.
The
results are listed in Table~\ref{tab:ions}. As for the uncharged
methane, the potential at the uncharged sodium site is slightly
positive. The decrease of $m_c$ with increasing charge is stronger than
linear and, correspondingly, the fluctuation $f_c$ increases slightly
with the charge. This indicates that a simple Gaussian model using an
expansion around the uncharged particle is of limited utility.

We use the information about the derivatives to calculate the free
energy of charging using polynomial fits. The results for the sodium
ion using polynomials of degree 2, 4, and 6 are compiled in
Table~\ref{tab:Na}. Also included in Table~\ref{tab:Na} are results
obtained from TI, as described in section~\ref{sec:methods}. TI was
performed using Ewald summation and $N=8$, 16, 32, 64, 128, and 256
water molecules as well as using the GRF Coulomb interaction and $N=32$, 64,
and 128 water molecules. We observe excellent agreement of the
free-energy data from polynomial fits and TI, except for the $p_2$ fit
which cannot fully account for the increasing potential fluctuations
with increasing charge. The TI data of charging from 0 to $e$ and
uncharging from $e$ to 0 show variations of about 5~kJ~mol$^{-1}$.
Regarding the system-size dependence, Ewald summation gives accurate
results even for as few as $N=16$ water molecules. The GRF shows a more
pronounced system-size dependence with the $N=64$ data (cutoff
$r_c=0.62$~nm) being slightly too low. These results indicate that the
free energy of charging is unexpectedly insensitive to the system size
if the electrostatic interactions are treated appropriately. In
particular, it is important to apply the correct finite-size
corrections. For Ewald summation with $N=16$, for instance, the
finite-size correction accounts for about 60 percent of the free energy.
Without the self terms the Ewald results for $N=256$ and $N=16$ differ
by about 63~kJ~mol$^{-1}$; with the self terms included the difference
is only 5~kJ~mol$^{-1}$.

Table~\ref{tab:dF_ions} lists the results of polynomial fits of the
free energy to the derivative data for the other ions studied (K$^+$,
Ca$^{2+}$, F$^-$, Cl$^-$, and Br$^-$). Also included are results of TI
calculations using Ewald summation and $N=128$ water molecules. Except
for the polynomial fit of degree 2, we obtain consistent results from
the derivative data and TI. The $p_2$ results are always somewhat too
negative but this is more apparent for the negative ions. The two TI
data per ion typically bracket the $p_4$ and $p_6$ results for the
free energy.

Interestingly, there is no simple trend for the free energy of charging
of monovalent cations with the ion size (as measured by $\sigma$ of the
LJ interaction). The positive ions Na$^+$ and K$^+$ as well as the
negative ions F$^-$, Cl$^-$, and Br$^{-}$ show the expected decrease of
$\Delta \mu^{ex}$ with increasing $\sigma$. However, only the
negatively charged methane-like LJ particle Me$^-$ fits into this
ordering. The positively charged Me$^+$ has a less negative $\Delta
\mu^{ex}$ than K$^+$, even though the van der Waals diameter $\sigma$ of
K$^+$ is considerably larger. However, the LJ interaction of the
K$^+$ ion is more shallow than that of Me$^+$ with the LJ $\epsilon$
values differing by a factor of about 150.

We also calculated the excess chemical potential of inserting
uncharged LJ particles in SPC water of density $\rho=33.33$~nm$^{-3}$
at temperature $T=298$~K. This was done using test-particle
insertion.\cite{Widom:82,Pohorille:90,Pratt:92,Pratt:93,Guillot:91,%
Guillot:93:a,Guillot:93:b,Madan:94,Forsman:94,Beutler:95:a} 5000 SPC
water configurations were used of a simulation run extending over
500\,000 Monte Carlo passes. The simulation was performed using
$N=256$ water molecules and GRF Coulomb interaction with a cutoff of
$r_c=0.9$~nm. We calculated $\langle\exp(-\beta u)\rangle$ using 100
test particles per configuration, where $u$ is the interaction energy
of a LJ test particle with the water molecules. For the LJ
interaction, a spherical cutoff distance of $L/2=0.9865$~nm was
used. A cutoff correction for the $r^{-6}$ term was applied, assuming
homogeneous water density beyond the cutoff. The excess chemical
potential is calculated as
\begin{eqnarray}
\mu^{ex} & = & - k_{\rm B}T \ln \langle\exp(-\beta u)\rangle~.
\end{eqnarray}
Results are listed in Table~\ref{tab:muex}. We find positive values
for $\mu^{ex}$ between 9 and 25~kJ~mol$^{-1}$, favoring the gaseous
state. Adding $\mu^{ex}$ to the free energy of charging, we obtain
single-ion free energies of hydration.

Experimental data for single-ion free energies of hydration have been
compiled by, for instance, Friedman and Krishnan,\cite{Friedman:73}
Conway,\cite{Conway:78} and most recently Marcus.\cite{Marcus:91:a}
The first two references report values for the standard molar Gibbs free
energy $\Delta G^0$, {\em i.e.}, for a hypothetical transfer from a
1~atm gas state to a 1~mol/l solution. Marcus lists values for $\Delta
G^*$ which is the Gibbs free energy of bringing an ion from an empty box
into solution. The theoretical calculations determine the excess free
energy of hydration, {\em i.e.}, the transfer from an ideal gas of given
density to solution of equivalent solute density. This process
corresponds to that used by Marcus, so that $\Delta G^*$ is the
experimental equivalent of the theoretical free energy that we have
referred to as $\mu^{ex}$ disregarding volume contributions.
Because Marcus used $\Delta G^*$ for the
experimental free energies of hydration we will retain that notation
here for those quantities. To convert from $\Delta G^0$ to $\Delta
G^*$, requires adjustment for the differences in standard states: we
add to $\Delta G^0$ the free energy of an ideal gas going from a
pressure $p_0$ corresponding to a density of 1~mol/l to a pressure
$p_1=1$~atm, which is $k_{\rm B}T \ln ( p_1 / p_0 )$, {\em i.e.},
$\Delta G^*=\Delta G^0 - 7.92$~kJ~mol$^{-1}$.\cite{conversion}
Another correction accounts for differing values for the reference ion
H$^+$. We take the most recent value by Marcus\cite{Marcus:91:a}
$\Delta G^*[\mbox{H}^+] = -1050\pm 6$~kJ~mol$^{-1}$ and adjust the
other values [$-1098$ (ref \onlinecite{Friedman:73}) and $-1074\pm
17$~kJ~mol$^{-1}$ (ref \onlinecite{Conway:78})] accordingly.

Results for the calculated free energy of ionic hydration $\mu^{ex}=
\mu^{ex}(q=0)+\Delta \mu^{ex}(0\rightarrow q)$ and the experimental
values $\Delta G^*$ are compiled in Table~\ref{tab:Fhyd}. For the
calculated values we use those obtained from a fit of a sixth-order
polynomial $p_6$ to the derivative data, as listed in
Table~\ref{tab:dF_ions}. The experimental data were adjusted as
described above. The experimental data for cations show little
variation between the three sources. However, the anion data vary by
as much as 70~kJ~mol$^{-1}$, with the Conway data\cite{Conway:78}
bracketed by the those of refs
\onlinecite{Marcus:91:a} and \onlinecite{Friedman:73}, but generally
closer to the data of Marcus.\cite{Marcus:91:a}

The calculated free energy data for cations do not show a clear trend.
The results for Na$^+$ and K$^+$ are too low and too high by about 10
percent, respectively. The hydration free energy of Ca$^{2+}$ is too
high by about 15 percent. The anions on the other hand show a clear
tendency with the magnitudes of the calculated free energies generally
being too large. The relative errors are 26, 10, and 15 percent for
F$^-$, Cl$^-$, and Br$^-$, respectively, compared to the data of Marcus.
These significantly too negative values of the hydration free energy of
anions might be a consequence of the unprotected hydrogen atoms in the
water-ion interaction model used. The positively charged hydrogen atom
can penetrate the LJ shell of the ions without a direct energetic
penalty. The interaction with the negative point charge at the center
of the ion strongly binds the water molecule, resulting in a large
enthalpic contribution to the free energy of hydration. But also
effects of non-additive interactions might play a considerable
role.\cite{Xantheas:92}

Also included in Table~\ref{tab:Fhyd} are computer simulation results
by Straatsma and Berendsen.\cite{Straatsma:88} These authors used
thermodynamic integration in conjunction with isothermal-isobaric
molecular dynamics simulations to compute hydration free energies of
ions. The interaction potentials used here are identical with those of
Straatsma and Berendsen, except for the treatment of the electrostatic
interactions. We used Ewald summation; whereas Straatsma and Berendsen
used a spherical cutoff and a Born-type correction for finite-size
effects. These authors (and others\cite{Marrone:93}) argue that the
application of a Born-type correction is rather crude, approximating
the solvent molecules beyond the cutoff by a dielectric continuum.
Nevertheless, in the absence of a better alternative it has been
widely adopted. Migliore {\em et al.}\cite{Migliore:88} calculated the
free energy of ionic hydration based on a perturbation formula from
Monte Carlo simulations using MCY water and {\em ab initio} ion-water
potentials. These authors also used a spherical cutoff.
Table~\ref{tab:Fhyd} includes the results of Migliore {\em et al.},
who did not apply a finite-size correction.

Qualitatively, our free-energy data agree with those of Straatsma and
Berendsen\cite{Straatsma:88} and Migliore {\em et
al.}\cite{Migliore:88}\ \ We observe the same ordering of the free
energies with respect to ion size. The quantitative agreement is
however poor. Our values for the cations Na$^+$ and K$^+$ are closer
to the experimental data of Marcus. The cation free energies of
Straatsma and Berendsen (with Born correction) are consistently more
negative than those of our calculations. On the other hand, our anion
free energies are significantly more negative than those of Straatsma
and Berendsen as well as of Migliore {\em et al.}\ \ The results of
Straatsma and Berendsen for Cl$^-$ and Br$^-$ are somewhat closer to
the experimental data of Marcus when the Born correction is included.
Without the correction they are significantly too high. The most
pronounced discrepancies between the anion data of Straatsma and
Berendsen\cite{Straatsma:88} and ours are those of the fluoride ion,
with our $\mu^{ex}$ values being lower by 83~kJ~mol$^{-1}$. This is
somewhat surprising since Straatsma and Berendsen used the same
parameters for the water-water and water-ion interactions. The
difference can be a consequence of using different ensembles (NVT
versus quasi-NPT); or, more likely, it is caused by the different
treatment of the electrostatic interactions (Ewald versus spherical
cutoff).

The fluoride ion also shows the largest relative deviations from the
experimental results. To further investigate these discrepancies, we
have studied the energetics of clusters of different size formed by a
single fluoride ion and water. We have performed Monte Carlo
simulations using one F$^-$ ion that nucleates $N=4$, 8, 12, 16, 32,
64, 128, and 256 water molecules at 298~K, as described in
section~\ref{sec:methods}. We calculated the interaction energy $u_s$
of the fluoride ion with the SPC water molecules over 50\,000 passes.
Figure~\ref{fig:clust} shows the differences $\Delta u_s = u_s -
u_{s,{\rm EW}}$ with respect to the bulk simulation using $N=128$
water molecules and Ewald summation as a function of $N^{-1/3}$.
$\Delta u_s$ can be fitted to a third-order polynomial in $N^{-1/3}$
over the whole range of system sizes. Extrapolation to
$N\rightarrow\infty$ yields a limit for $\Delta u_s$ close to zero.
(However, the nontrivial dependence on $N^{-1/3}$ limits the accuracy
of the extrapolation.) The result obtained from Ewald summation,
$u_{s,{\rm EW}}=-1077\pm 4$~kJ~mol$^{-1}$, also agrees with the value
$u_s=-1075$~kJ~mol$^{-1}$ obtained from integrating the pair
correlation functions of the bulk simulation using $\varphi(r)=1/r$ in
eq~\ref{eq:gint}, adding the LJ contributions, and applying the
finite-size correction $-e^2 \xi_{1/r}$. The integration shows that
the LJ contributions are strongly repulsive ($\approx
90~$kJ~mol$^{-1}$) but compensated by large electrostatic
interactions.

The value for the solvation energy reported by Straatsma and Berendsen,
$u_s = -823$~kJ~mol$^{-1}$, is however considerably smaller. The
observed differences in $u_s$ of about 150~kJ~mol$^{-1}$ agree in
magnitude and sign with those of the free energies (83~kJ~mol$^{-1}$).
If we truncate the integration of $1/r$ weighted with the pair correlation
functions obtained from Ewald summation at $R=0.9$~nm (which is the
cutoff Straatsma and Berendsen used) and do not apply a finite-size
correction, we obtain a value of $-867$~kJ~mol$^{-1}$ in much closer
agreement with Straatsma and Berendsen's. This indeed indicates that
the treatment of the electrostatic interactions (Ewald summation versus
spherical cutoff) is the major source of the discrepancy.

Also included in Figure~\ref{fig:clust} are the results for the mean and
the variance of the electrostatic potential at the ion site.
Figure~\ref{fig:clust} shows differences with respect to the bulk value.
The differences of the mean values $\Delta\langle\phi\rangle$ closely
follow the solvation-energy differences $\Delta u_s$ and can also
be fitted to a third-order polynomial in $N^{-1/3}$. The differences of
the fluctuation $\Delta\langle\Delta\phi^2\rangle$ depend linearly on
$N^{-1/3}$ for $N$ between 8 and 256. Both fitted curves extrapolate
to approximately 0, indicating that the calculated bulk values are
the correct limits for $N\rightarrow\infty$.

{}From the cluster-size dependence of the solvation energy and the mean
and variance of the electrostatic potential, as well as the results for
Me and Na$^+$, we conclude that the use of periodic boundary conditions
in conjunction with Ewald-summation (or reaction-field) electrostatics
closely approximates the correct bulk behavior of the system; however,
to get correct energetics it is important to include the self-interactions
in the Coulomb energy.

\section{Conclusions}
We have shown that free energies can be accurately
calculated from equilibrium simulations by extracting derivative
information with respect to a coupling parameter. We have studied the
free energy of charging ions in water, which accounts for most of the
free energy of ionic solvation for typical ion sizes. The choice of the
ionic charge as coupling parameter results in free-energy expressions
involving cumulants of the electrostatic potential $\phi$ at the charge
sites. We find that the statistics of $\phi$ are approximately
Gaussian. This means that only the first and second moment of the
distribution can be calculated accurately, with higher moments dominated
by the poorly sampled tails. Correspondingly, only information about
the first and second derivative of the free energy can be calculated
accurately for any given charge state. The information for different
charge states ({\em e.g.}, uncharged and fully charged) can then be combined
using interpolation or polynomial fitting.

We have studied a methane-like Lennard-Jones particle in SPC water. We
observe two almost Gaussian regimes separated by $q=0$ with different
characteristics. Negative ions are more stably solvated compared to
positive ions of equal size, in agreement with the experimental
data.\cite{Latimer:39} The system shows further asymmetry, since the
average electrostatic potential at the position of the uncharged
particle is positive. This means that increasing the ion charge first
costs energy. We relate these asymmetries of the energetics (lower free
energy of negative ions, positive potential) to the structural asymmetry
of the water molecule. The hydrogen atoms can penetrate the ionic van
der Waals shell, whereas the oxygen atom is better protected. For the
uncharged particle, this results in a net positive potential; and the
point charge at the center of negative ions exerts strong
electrostatic interactions with the tightly bound hydrogen of water.

However, particularly for small anions this effect might by exaggerated
by the interaction potentials used. This potential model does not give
a protective van der Waals sphere to the charge on the hydrogen atom.
In principle, this is a fundamental difficulty, but in computer
simulations, the heights of energetic barriers usually exclude the
singularity. The development of interaction potentials for anion-water
interactions nevertheless has to account for these problems. The strong
interactions with the hydrogens ``pull'' the water closer and the first
maxima of the ion-oxygen pair correlation function is already in the
strongly repulsive region, reducing the effective ion radius.

We have also studied the charging of sodium, potassium, calcium,
fluoride, chloride, and bromide ions. The agreement with the
available experimental data for solvation free energies is only
qualitative, reproducing the trends with ionic size. The quantitative
data are not in satisfactory agreement with the experimental results,
even conceding quite substantial discrepancies between different
compilations of the experimental data for certain ions. We observe
typical errors of about 10 to 15 percent for the free energies of
ionic solvation compared to the experimental data of
Marcus.\cite{Marcus:91:a} This clearly indicates the further need to
develop quantitatively reliable descriptions of ion-water
interactions.

However, to allow for valid comparisons of data obtained from
computer simulations with experimental results, it is crucial to
eliminate systematic errors in the simulation methods. An important
part of this study was devoted to analyzing the effect of finite system
sizes on the free energy of charging. We could clearly establish that
Ewald summation (and, similarly, the generalized reaction-field method)
accounts for finite-size effects by adding a term that corrects
for self-interactions. We showed that even for systems with only $N=16$
water molecules it is possible to obtain accurate estimates of the
solvation free energy of the sodium ion. For typical system sizes of a
few hundred water molecules, these finite-size corrections are
substantial in magnitude. Neglecting them yields results of little
quantitative validity.

\section*{Acknowledgments}
G. H. wishes to thank M. Neumann and D. M. Soumpasis for many
stimulating discussions. This work was funded by the Department of
Energy (U.S.).

\begin{figure}
\caption{Probability distributions $P(e\phi)$ of the electrostatic
energy $e\phi$ at the site of a methane-like ion Me with charge $q=e$
from Ewald summation with $N=256$ ($\diamond$, \mbox{------}), $N=128$
($\Box$, \mbox{--}~$\cdot$~\mbox{--}), and GRF with $N=256$ water
molecules ($+$, \mbox{--}~\mbox{--}~\mbox{--}), respectively. The
lines are Gaussian distributions. Also shown are Gaussian
distributions corrected for finite-size effects, which are peaked near
$e\phi=-550$~kJ~mol$^{-1}$; they agree closely in position and
variance.}
\label{fig:histo}
\end{figure}

\begin{figure}
\caption{The average electrostatic potential $\phi$ at the site of the
negatively charged ion Me ($q=-e$) calculated from the pair correlations
of a Monte Carlo simulation using Ewald summation and $N=256$ water
molecules. The results of the integration using the GRF interaction
with cutoff $r_c=L/2$ and the bare Coulomb interaction $1/r$ are shown
with long- and short-dashed lines, respectively. Finite-size
corrections are added as constants. The Ewald-summation result is
shown as a reference with a solid line.}
\label{fig:phi}
\end{figure}

\begin{figure}
\caption{The average electrostatic potential $m_c$ at the position of
the methane-like Lennard-Jones particle Me as a function of its charge
$q$. $m_c$ contains corrections for the finite system size. Results
are shown from Monte Carlo simulations using Ewald summation with
$N=256$ ($+$) and $N=128$ ($\times$) as well as GRF calculations with
$N=256$ water molecules ($\Box$). Statistical errors are smaller than
the size of the symbols. Also included are linear fits to the data
with $q<0$ and $q\geq 0$ (solid lines). The fit to the
$\tanh$-weighted model of two Gaussian distributions
(eq~\protect\ref{eq:tanh}) is shown with a dashed line.}
\label{fig:meth_m}
\end{figure}

\begin{figure}
\caption{The fluctuation of the electrostatic potential $f_c$ at the
position of a methane-like Lennard-Jones particle as a function of its
charge $q$. $f_c$ contains corrections for the finite system size.
Error bars indicate one estimated standard deviation of the data. For
further details see Figure~\protect\ref{fig:meth_m}.}
\label{fig:meth_f}
\end{figure}

\begin{figure}
\caption{The pair correlation functions $g_{\rm IO}$ (top panel)
and $g_{\rm IH}$ (bottom panel) of the Me ion with water oxygen
and hydrogen. The $g(r)$ curves are shifted vertically according to
the ionic charge by $q/e$, {\em i.e.}, by $1$ for $q=e$, $0.5$ for $q=0.5e$
etc.\ \ The $g(r)$ curves of Ewald summation and GRF simulations with
$N=256$ water molecules are shown with solid and dashed lines,
respectively.}
\label{fig:gr}
\end{figure}

\begin{figure}
\caption{The energetics of clusters of a fluoride ion and SPC water.
Results are shown for the interaction energy $u_s$ of the fluoride ion with
the water ($\diamond$), as well as the mean $\langle\phi\rangle$
($+$) and variance $\langle\Delta\phi^2\rangle$ ($\Box$) of the
electrostatic potential at the ion position. The figure shows
differences of these quantities with respect to the bulk values
calculated from Monte Carlo simulation of an $N=128$ water-molecule
system using Ewald summation: $\Delta u_s=u_s-u_{s,{\rm EW}},
\Delta\langle\phi\rangle = \langle\phi\rangle - \langle\phi\rangle_{\rm
EW}$, and $\Delta\langle\Delta\phi^2\rangle =
\langle\Delta\phi^2\rangle - \langle\Delta\phi^2\rangle_{\rm EW}$.
The lines are fitted curves as explained in the text. Error bars
indicate one standard deviation estimated from block averages.
The standard deviations of the bulk and cluster data were added.}
\label{fig:clust}
\end{figure}

\narrowtext
\begin{table}
\caption{Lennard-Jones parameters of the ion-water
interactions.\label{tab:LJ}}
\begin{tabular}{rrr}
Ion & $\epsilon/\mbox{(kJ mol$^{-1}$)}$ & $\sigma/\mbox{nm}$ \\ \hline
Na$^+$    & 0.200546 & 0.285000 \\
K$^+$     & 0.006070 & 0.452000 \\
Ca$^{2+}$ & 0.637972 & 0.317000 \\
F$^-$     & 0.553830 & 0.305000 \\
Cl$^-$    & 0.537866 & 0.375000 \\
Br$^-$    & 0.494464 & 0.383000 \\
Me    & 0.893228 & 0.344778 \\
\end{tabular}
\end{table}

\begin{table}
\caption{Results for the mean and the fluctuation of the potential
$\phi$ (with and without finite-size corrections) at the position of a
methane-like Lennard-Jones particle Me carrying a charge $q$.
``Coulomb'' refers to the treatment of the electrostatic interactions
(Ewald or GRF). $N$ is the number of water molecules. The mean and the
fluctuation are listed as $m=e\langle\phi\rangle$ and $f=\beta
e^2\langle(\phi-\langle\phi\rangle)^2\rangle$, both in units of
kJ~mol$^{-1}$. The corrected values are $m_c=m+qe\xi$ and
$f_c=f-e^2\xi$. The statistical errors of $m$ and $f$ are estimated
from block averages as approximately 4.0 and 30 kJ~mol$^{-1}$.
\label{tab:methane}}
\begin{tabular}{rrrrrrr}
$N$ & Coulomb & $q/e$ & $m$ & $f$ & $m_c$ & $f_c$ \\ \hline
$256 $& EW  &$ -1.00 $&$  670.3 $&$ 604 $&$  869.9 $&$ 804$\\
$128 $& EW  &$ -1.00 $&$  618.1 $&$ 664 $&$  869.2 $&$ 915$\\
$256 $& GRF &$ -1.00 $&$  520.5 $&$ 493 $&$  869.1 $&$ 842$\\
$128 $& EW  &$ -0.75 $&$  465.3 $&$ 730 $&$  653.6 $&$ 981$\\
$256 $& EW  &$ -0.50 $&$  324.0 $&$ 713 $&$  423.7 $&$ 913$\\
$128 $& EW  &$ -0.50 $&$  292.2 $&$ 698 $&$  417.8 $&$ 950$\\
$256 $& GRF &$ -0.50 $&$  242.2 $&$ 568 $&$  416.5 $&$ 917$\\
$128 $& EW  &$ -0.25 $&$  141.3 $&$ 529 $&$  204.0 $&$ 780$\\
$256 $& EW  &$  0.00 $&$   38.0 $&$ 367 $&$   38.0 $&$ 567$\\
$128 $& EW  &$  0.00 $&$   37.3 $&$ 341 $&$   37.3 $&$ 592$\\
$256 $& GRF &$  0.00 $&$   34.9 $&$ 239 $&$   34.9 $&$ 587$\\
$128 $& EW  &$  0.25 $&$  -40.7 $&$ 313 $&$ -103.5 $&$ 564$\\
$256 $& EW  &$  0.50 $&$ -143.1 $&$ 374 $&$ -242.9 $&$ 573$\\
$128 $& EW  &$  0.50 $&$ -118.1 $&$ 332 $&$ -243.6 $&$ 583$\\
$256 $& GRF &$  0.50 $&$  -74.4 $&$ 232 $&$ -248.7 $&$ 581$\\
$128 $& EW  &$  0.75 $&$ -205.9 $&$ 354 $&$ -394.2 $&$ 605$\\
$256 $& EW  &$  1.00 $&$ -348.8 $&$ 450 $&$ -548.3 $&$ 650$\\
$128 $& EW  &$  1.00 $&$ -298.9 $&$ 389 $&$ -550.0 $&$ 640$\\
$256 $& GRF &$  1.00 $&$ -202.2 $&$ 296 $&$ -550.9 $&$ 645$\\
\end{tabular}
\end{table}

\begin{table}
\caption{Free energy (in kJ mol$^{-1}$) of charging the methane-like
Lennard-Jones particle Me from 0 to $\pm e$. The free energy was
calculated from fitting to polynomials $p_n$ of degree $n$ and a
$\tanh$-weighted model of two Gaussian regimes
(eq~\protect\ref{eq:tanh}).\label{tab:df_meth}}
\begin{tabular}{rrr}
Function & $\Delta \mu^{ex}(0\rightarrow e)$ & $\Delta
\mu^{ex}(0\rightarrow-e)$
\\ \hline
$p_2$     &$ -246$ &$ -454$ \\
$p_4$     &$ -253$ &$ -431$ \\
$p_6$     &$ -250$ &$ -431$ \\
$p_8$     &$ -250$ &$ -431$ \\
$p_{10}$  &$ -250$ &$ -431$ \\
$\tanh$   &$ -250$ &$ -430$ \\
\end{tabular}
\end{table}

\begin{table}
\caption{Results for the mean $m_c$ and fluctuation $f_c$ of the
potential (with finite-size corrections included) at the position of
sodium, potassium, calcium, fluoride, chloride, and bromide ions at
different charge-states $q$. The data were calculated from Monte Carlo
simulations using $N=128$ water molecules and Ewald summation over
100\,000 passes. The mean and the fluctuation are listed as
$m_c=e(\langle\phi\rangle+q\xi)$ and $f_c=\beta
e^2\langle(\phi-\langle\phi\rangle)^2\rangle-e^2\xi$, both in
units of kJ~mol$^{-1}$. The statistical errors of $m_c$ and $f_c$ are
estimated from block averages to be approximately 4.0 and 30
kJ~mol$^{-1}$.\label{tab:ions}}
\begin{tabular}{rrrr}
Ion & $q/e$ & $m_c$ & $f_c$ \\ \hline
Na &$  0.00 $&$   39.0 $&$  891$\\
Na &$  0.50 $&$ -395.6 $&$  956$\\
Na &$  1.00 $&$ -885.1 $&$  970$\\
K  &$  0.00 $&$   38.6 $&$  682$\\
K  &$  0.50 $&$ -282.1 $&$  690$\\
K  &$  1.00 $&$ -662.6 $&$  772$\\
Ca &$  0.00 $&$   41.0 $&$  662$\\
Ca &$  1.00 $&$ -653.6 $&$  789$\\
Ca &$  2.00 $&$-1367.6 $&$  667$\\
F  &$  0.00 $&$   35.7 $&$  718$\\
F  &$ -0.50 $&$  587.6 $&$ 1381$\\
F  &$ -1.00 $&$ 1167.3 $&$  961$\\
Cl &$  0.00 $&$   36.2 $&$  550$\\
Cl &$ -0.50 $&$  378.2 $&$  819$\\
Cl &$ -1.00 $&$  794.1 $&$  773$\\
Br &$  0.00 $&$   37.3 $&$  545$\\
Br &$ -0.50 $&$  369.2 $&$  758$\\
Br &$ -1.00 $&$  772.7 $&$  773$\\
\end{tabular}
\end{table}

\begin{table}
\caption{Results for the free energy $\mu^{ex}$ of charging the sodium
cation from $q=0$ to $e$ in SPC water in units of kJ~mol$^{-1}$.
$\mu^{ex}$ includes the finite-size corrections which are listed as
$\mu^{ex}_{\rm self}$. The free energies were calculated from
polynomial fits to the derivative data of Table~\protect\ref{tab:ions}
(polynomials $p_n$ of degree $n$). Also included are results of
thermodynamic integration (TI). Linear charging paths from 0 to $e$
and from $e$ to 0 are denoted by upward ($\uparrow$) and downward
($\downarrow$) arrows, respectively.
Ewald (EW) and generalized reaction-field (GRF)
interactions were used for the charges. \label{tab:Na}}
\begin{tabular}{rrrrr}
Method & Coulomb & $N$ & $\mu^{ex}_{\rm self}$ & $\mu^{ex}$ \\ \hline
$p_2$          & EW  & 128 &$ -126 $&$ -415 $\\
$p_4$          & EW  & 128 &$ -126 $&$ -407 $\\
$p_6$          & EW  & 128 &$ -126 $&$ -407 $\\
TI$\uparrow$   & EW  & 256 &$ -100 $&$ -404 $\\
TI$\downarrow$ & EW  & 256 &$ -100 $&$ -406 $\\
TI$\uparrow$   & EW  & 128 &$ -126 $&$ -402 $\\
TI$\downarrow$ & EW  & 128 &$ -126 $&$ -407 $\\
TI$\uparrow$   & EW  &  64 &$ -158 $&$ -407 $\\
TI$\downarrow$ & EW  &  64 &$ -158 $&$ -406 $\\
TI$\uparrow$   & EW  &  32 &$ -198 $&$ -403 $\\
TI$\downarrow$ & EW  &  32 &$ -198 $&$ -407 $\\
TI$\uparrow$   & EW  &  16 &$ -247 $&$ -409 $\\
TI$\downarrow$ & EW  &  16 &$ -247 $&$ -411 $\\
TI$\uparrow$   & EW  &   8 &$ -305 $&$ -419 $\\
TI$\downarrow$ & EW  &   8 &$ -305 $&$ -425 $\\
TI$\uparrow$   & GRF & 128 &$ -219 $&$ -401 $\\
TI$\downarrow$ & GRF & 128 &$ -219 $&$ -406 $\\
TI$\uparrow$   & GRF &  64 &$ -276 $&$ -408 $\\
TI$\downarrow$ & GRF &  64 &$ -276 $&$ -411 $\\
TI$\uparrow$   & GRF &  32 &$ -346 $&$ -419 $\\
TI$\downarrow$ & GRF &  32 &$ -346 $&$ -424$
\end{tabular}
\end{table}

\begin{table}
\caption{Results for the free energy $\mu^{ex}$ of charging the
potassium, calcium, fluoride, chloride, and bromide
ions from $q=0$ to $\pm e, 2e$ in SPC water in units of
kJ~mol$^{-1}$. $\mu^{ex}$ includes finite-size corrections.
Details as in Table~\protect\ref{tab:Na}.}
\label{tab:dF_ions}
\begin{tabular}{rrrrrr}
Ion & $p_2$ & $p_4$ & $p_6$ & TI$\uparrow$ & TI$\downarrow$ \\ \hline
K$^+$     &$  -297 $&$  -293 $&$  -295 $&$  -291 $&$  -294$\\
Ca$^{2+}$ &$ -1317 $&$ -1315 $&$ -1316 $&$ -1311 $&$ -1327$\\
F$^-$     &$  -594 $&$  -590 $&$  -590 $&$  -590 $&$  -594$\\
Cl$^-$    &$  -401 $&$  -392 $&$  -392 $&$  -389 $&$  -394$\\
Br$^-$    &$  -393 $&$  -382 $&$  -382 $&$  -379 $&$  -382$
\end{tabular}
\end{table}

\begin{table}
\caption{Results for the excess chemical potential $\mu^{ex}$ (in
kJ~mol$^{-1}$) of transferring an uncharged LJ particle from ideal gas
into SPC water. The LJ parameters are those of
Table~\protect\ref{tab:LJ}. Errors are estimated from block averages.}
\label{tab:muex}
\begin{tabular}{rr}
LJ-Particle & $\mu^{ex}$ \\ \hline
Na       &  9.2(1) \\
K        & 23.7(5) \\
Ca       & 10.2(3) \\
F        &  9.7(2) \\
Cl       &   21(3) \\
Br       &   24(3) \\
Me       & 10.2(9) \\
\end{tabular}
\end{table}

\begin{table}
\caption{Results for the calculated free energy of ionic hydration (in
kJ~mol$^{-1}$) compared with experimental data. The experimental data
were adjusted to give $\Delta G^*=-1050$~kJ~mol$^{-1}$ for H$^+$, as
used by Marcus.\protect\cite{Marcus:91:a} Also included are computer
simulation results by Straatsma and
Berendsen\protect\cite{Straatsma:88} and Migliore {\em et
al.}\protect\cite{Migliore:88}}
\label{tab:Fhyd}
\begin{tabular}{rrrrrrrr}
Ion & $\mu^{ex}$ &
$\Delta G^*$ \tablenote{Experimental data of
Marcus.\protect\cite{Marcus:91:a}}&
$\Delta G^*$ \tablenote{Experimental data of Friedman and
Krishnan.\protect\cite{Friedman:73}}&
$\Delta G^*$ \tablenote{Experimental data of
Conway.\protect\cite{Conway:78}}&
$\mu^{ex}$ \tablenote{Computer simulation data of Straatsma and
Berendsen calculated using molecular dynamics of $N=216$ water
molecules.\protect\cite{Straatsma:88} The results contain a
Born-type correction applied by the authors to their raw data.}&
$\mu^{ex}$ \tablenote{Computer simulation data of Straatsma and
Berendsen without Born correction.\protect\cite{Straatsma:88}}&
$\mu^{ex}$ \tablenote{Computer simulation data of Migliore {\em et
al.} calculated using molecular dynamics of $N=342$ water
molecules.\protect\cite{Migliore:88}}
\\ \hline
Na$^+$     &$ -398$&$ -365$&$ -371$&$ -372$&$ -508$&$ -431$&$-459$\\
K$^+$      &$ -271$&$ -295$&$ -298$&$ -298$&$ -425$&$ -349$&$-321$\\
Ca$^{2+}$  &$-1306$&$-1505$&$-1553$&  --- &$-1623$&$-1394$& --- \\
F$^-$      &$ -580$&$ -465$&$ -394$&$ -441$&$ -497$&$ -421$&$-418$\\
Cl$^-$     &$ -371$&$ -340$&$ -277$&$ -324$&$ -315$&$ -239$&$-237$\\
Br$^-$     &$ -358$&$ -315$&$ -263$&$ -310$&$ -304$&$ -228$& --- \\
\end{tabular}
\end{table}


\begin{thebibliography}{10}
\bibitem{minimg}
Refs \protect\onlinecite{Nijboer:88} and
\protect\onlinecite{Sorensen:91} present rather lengthy calculations
of the numerical value of $\xi_{1/r}$. We note that the integral of
$1/r$ over a cubic volume can easily be obtained by summing the
results of partial integration in each of the 3 dimensions,
\begin{eqnarray} -L\xi_{1/r} &=& \int_{-1/2}^{1/2}dx \int_{-1/2}^{1/2}dy
\int_{-1/2}^{1/2}dz (x^2+y^2+z^2)^{-1/2}\nonumber\\ & = & \frac{3}{2}
\int_{-1/2}^{1/2}dx \int_{-1/2}^{1/2}dy (x^2+y^2+1/4)^{-1/2} \nonumber\\ &=&
\frac{3}{2}
\int_{-1/2}^{1/2}dx \left\{\ln\left[\left(4x^2+2\right)^{1/2}+1\right]-
\ln\left[\left(4x^2+2\right)^{1/2}-1\right]\right\}\nonumber\\ &=&
3\ln\left(3^{1/2}+2\right)-\pi/2~.\nonumber \end{eqnarray}
\bibitem{conversion}
Interestingly, Marcus\protect\cite{Marcus:94:a,Marcus:91:a} finds the
correction to be of opposite sign. Straatsma and
Berendsen\cite{Straatsma:88} however determine a correction identical
to ours. Assuming the tabulated values of Marcus to be correct, this
does however not affect the following discussion.
\end{thebibliography}
\end{document}